\documentclass[conference, 10pt]{IEEEtran}
\IEEEoverridecommandlockouts
\usepackage{cite}
\usepackage{amsmath,amssymb,amsfonts,float,graphicx,steinmetz,bm,subfigure,multicol,multirow,url}
\usepackage{textcomp}


\begin{document}             
\title{
RIS-Assisted MISO Communication: Optimal Beamformers and Performance Analysis
\thanks{This work was supported by the HKTIIT (grant number HKTIIT16EG01). It was also supported in part by the Hong Kong PhD Fellowship Scheme (PF17-00157).
}
\thanks{This work has been accepted for presentation at IEEE GlobeCom-2020.
}
}


\author
{
\IEEEauthorblockN{Neel Kanth Kundu, {\em Student Member, IEEE}, and
Matthew R. McKay, {\em Senior Member, IEEE}}\\
\IEEEauthorblockA{\textit{ECE Department,}
The Hong Kong University of Science and Technology,\\
Clear Water Bay, Kowloon, Hong Kong \\
Email: nkkundu@connect.ust.hk, m.mckay@ust.hk}
}



\maketitle

\begin{abstract}
We study a multiple-input single-output (MISO) communication system assisted by a reconfigurable intelligent surface (RIS). A base station (BS) having multiple antennas is assumed to be communicating to a single-antenna user equipment (UE), with the help of a RIS. We assume that the system operates in an environment with line-of-sight (LoS) between the BS and RIS, whereas the RIS-UE link experiences Rayleigh fading. We present a closed form expression for the optimal active and passive beamforming vectors at the BS and RIS respectively, which require only the knowledge of the cascaded (BS-RIS-UE) channel at the BS. Then, by characterizing the statistical properties of the received SNR at the UE, we apply them to derive analytical approximations for different system performance measures, including the outage probability, average achievable rate and average symbol error probability (SEP). Our results, in general, demonstrate that the gain due to RIS can be substantial, and can be significantly greater than the gains reaped by using multiple BS antennas. 
\end{abstract}


\IEEEpeerreviewmaketitle
\section{Introduction}
To meet the high data rate, high reliability, and energy efficiency requirements of the emerging sixth-generation (6G) wireless communication systems, new physical layer technologies are being investigated \cite{dang2020should,kundu2020signal}. Among other potential technologies, RIS is envisioned as a new physical layer technology that can provide spectral and energy efficiency gains reminiscent of massive multiple-input multiple-output (MIMO), but with much fewer antennas at the BS \cite{huang2019reconfigurable,risdi2019smart}. Since RIS is composed of almost passive elements that can intelligently control the propagation of the impinging electromagnetic waves, they are energy efficient alternatives to massive MIMO. RIS has started to attract much attention from communication engineers, with multiple works focusing on optimizing the active and passive beamforming vectors at the transmitter and RIS respectively, with the aim of maximizing the spectral efficiency \cite{wu2019intelligent,yu2019miso,yang2019intelligent} or energy efficiency \cite{huang2019reconfigurable}. In these prior works, the beamforming vectors are obtained by solving complex optimization problems using numerical optimization tools (e.g., CVX \cite{cvx}) or iterative algorithms. The lack of a closed-form solution for these beamformers also prohibits analytical performance evaluation.


Generally, research on RIS is still in its infancy and the fundamental performance limits of this new technology are not fully understood; though, some contributions along this line are starting to emerge. Notably, for a RIS-assisted single-input single-output (SISO) system, system-level performance analyses were presented in some initial works \cite{basar2019wireless,kudathanthirige2020performance,badiu2019communication}. RIS-assisted MISO communication systems were studied in \cite{zhou2020spectral,wang2019intelligent}, which designed joint active and passive beamformers, and characterized the average received SNR. In \cite{zhou2020spectral},  both the BS-RIS and RIS-UE links were assumed deterministic LoS, whereas in \cite{wang2019intelligent}, only the BS-RIS link was assumed LoS, with the RIS-UE and BS-UE links assumed to be Rayleigh faded.



In this paper, we study a RIS-assisted MISO system for which the BS-RIS link is LoS and, like \cite{wang2019intelligent}, the RIS-UE link is subjected to Rayleigh fading. We derive analytical expressions for the optimal active and passive beamforming vectors at the BS and RIS that maximize the received SNR at the UE. These optimal beamforming vectors depend only on the cascaded channel, which can be  estimated at the BS. This is in contrast to the beamforming design proposed in \cite{zhou2020spectral}, which requires knowledge of the BS-RIS link and the RIS-UE link separately, which may be difficult to estimate in practice due to the passive nature of the RIS elements. Further, we present a performance analysis of the system, deriving closed form approximations for the outage probability, average achievable rate and average SEP. Asymptotic analysis of the outage probability reveals that the diversity order of the system depends only on the number of RIS elements, but not on the number of BS antennas. Our simulation results show that increasing the number of RIS elements leads to a substantial performance improvement, often exceeding the gain achieved by increasing the number of BS antennas. 


\section{System Model and Optimal Beamformers}
\subsection{System Model}
We consider a system where a BS having $M$ antennas is communicating to a single antenna UE with the help of a RIS which consists of $K$ passive elements. We assume that the direct channel between the UE and BS is very weak due to excessive blockage from trees, buildings etc, and only the BS-RIS-UE link can be used for communication \cite{zhou2020spectral,kudathanthirige2020performance,badiu2019communication}. We assume a flat fading scenario where the channel between the BS and RIS, $\bm{H} \in {\mathbb C}^{K \times M}$ consists of only LoS component and the channel between RIS and UE, $\bm{h}^T \in {\mathbb C}^{1 \times K}$ experiences Rayleigh fading, i.e., $\bm{h} \sim {\mathcal CN} (\bm{0}, \bm{I}_K)$. The received signal at the UE $y \in {\mathbb C}$ is given by
\begin{equation}
    y = \sqrt{P_{{\rm tx}}} \bm{h}^T {\rm diag}(\bm{\phi})\bm{H} \bm{w} x + n
    \label{e2.1}
\end{equation}
where $x$ is an information bearing symbol with ${\mathbb E}[|x|^2]=1$, $P_{{\rm tx}}$ denotes transmit power at the BS, and $n \sim {\mathcal CN} (0, \sigma^2)$ represents noise at the UE. Further, $\bm{w} \in {\mathbb C}^{M} $ is the (active) beamforming vector at the BS, and  $\bm{\phi} = [e^{j \theta_{1}}, \ldots, e^{j \theta_{K}} ]^{T} \in {\mathbb C}^{K}$ denotes a phase shift vector of the RIS, where $\theta_{k} \in [0, 2\pi]$. We assume that the BS and RIS are composed of uniform square planar arrays (USPA) such that the deterministic LoS channel between them can be expressed as 
\begin{equation}
    \bm{H} = \bm{a}_{K}\left(\psi_{r}^{a},\psi_{r}^{e}\right)\bm{a}_{M}^{H}\left(\psi_{t}^{a},\psi_{t}^{e}\right)
    \label{e2.2}
\end{equation}
where $\psi_{r}^{a}$ and $ (\psi_{r}^{e})$ are the azimuth and elevation angle of arrival (AoA) at the RIS respectively, $\psi_{t}^{a}$ and $ (\psi_{t}^{e})$ are the azimuth and elevation angle of departure (AoD) at the BS respectively. Further, $ \bm{a}_{K}\left(\psi_{r}^{a},\psi_{r}^{e}\right)$ , $ \bm{a}_{M}\left(\psi_{t}^{a},\psi_{t}^{e}\right)$ are the array response vectors at the RIS and BS. The array response vector $\bm{a}_{\ell}\left(\theta^{a},\theta^{e}\right)$ of a $\sqrt{\ell} \times \sqrt{\ell}$ USPA is given by \cite{zhou2020spectral} 
\begin{equation}
\begin{split}
 \bm{a}_{\ell}\left(\theta^{a},\theta^{e}\right) = [1,\ldots,e^{j2\pi \frac{d}{\lambda}(x\sin{\theta^a}\sin{\theta^e}+y\cos{\theta^e}))}, \\ \ldots, e^{j2\pi \frac{d}{\lambda}((\sqrt{\ell}-1)\sin{\theta^a}\sin{\theta^e}+(\sqrt{\ell}-1)\cos{\theta^e}))}]   
\end{split}
\label{e2.3}
\end{equation}
where $d$ is the inter-element spacing at the USPA, $\lambda$ is the wavelength of the signal, and $0 \leq x,y \leq \sqrt{\ell}-1$ are the element indices. It is convenient to rewrite ($\ref{e2.1}$) as
\begin{equation}
 y = \sqrt{P_{{\rm tx}}}\bm{\phi}^T \bm{V} \bm{w} x + n
 \label{e2.4}
\end{equation}
where $\bm{V}={\rm diag}(\bm{h}^T) \bm{H}$ is the effective cascaded channel between the BS and the UE and is assumed to be perfectly known at the BS. In practice, this cascaded channel may be estimated at the BS using existing channel estimation protocols (e.g., \cite{mishra2019channel,jensen2019optimal}).

\subsection{Optimum Active and Passive Beamforming Vectors}
The received SNR at the UE is given by
\begin{equation}
    \gamma = \Bar{\gamma} |\bm{\phi}^T \bm{V} \bm{w}|^2
    \label{e2.5}
\end{equation}
where $ \Bar{\gamma} = \frac{P_{{\rm tx}} }{\sigma^2}$ is the average transmit SNR.
It is known that for a fixed $\bm{\phi}$, the optimal beamforming vector at the BS that maximizes the received SNR (\ref{e2.5}) is given by
\begin{equation}
    \bm{w}_{{\rm opt}} =\frac{\left(\bm{\phi}^T \bm{V} \right)^H}{||\bm{\phi}^T \bm{V}||} \;.
    \label{e2.6}
\end{equation}
The received SNR at the UE is then
\begin{equation}
    \gamma = \Bar{\gamma} ||\bm{\phi}^T \bm{V}||^2 \;.
    \label{e2.7}
\end{equation}
We want to find the optimum passive beamforming vector $\bm{\phi}_{{\rm opt}}$ that maximizes the received SNR in (\ref{e2.7}); that is,
\begin{equation}
    \begin{split}
   \bm{\phi}_{{\rm opt}} =\; & \underset{ \bm{\phi}} {{\rm arg\, max}} \quad \bm{\phi}^{H}\bm{R} \, \bm{\phi} \\ &
    {\rm s.t} \quad \quad |\bm{\phi}_i|=1, \; \forall \; i= 1,\ldots,K
    \end{split}
    \label{e2.8}
\end{equation}
where $\bm{R}= \bm{V}^{*} \bm{V}^{T}$.  This optimization problem is non-convex due to the unit modulus constraint on the elements of $\bm{\phi}$. The problem (\ref{e2.8}) is recognized as a uni-modular quadratic program (UQP). Such problems arise in applications including radar waveform design, phase recovery, and active sensing \cite{soltanalian2014designing,zhang2006complex}. In general, UQPs are NP-hard problems, and several semi-definite programming relaxation and generalized power method based algorithms have been proposed to approximately solve them \cite{boumal2016nonconvex,kyrillidis2011rank,zhang2006complex}. However, under certain conditions on $\bm{R}$, an analytical solution for the global optimizer exists; see \cite[~Theorem 1]{soltanalian2014designing}. For our problem, the channel matrix $\bm{H}$ is rank 1, and consequently so are the matrices $\bm{V}$ and $\bm{R}$. Introduce the eigen decomposition $\bm{R} = \bm{U}\bm{\Lambda} \bm{U}^H$. Then the objective in (\ref{e2.8}) can be expressed as
\begin{equation}
    \bm{\phi}^{H}\bm{R} \, \bm{\phi} = \bm{\phi}^{H}\bm{U}\bm{\Lambda} \bm{U}^H \bm{\phi} = \lambda_1 |\bm{u}_{1}^{H}\bm{\phi}|^2 \;,
    \label{e2.9}
\end{equation}
where $\bm{u}_1$ is the eigenvector of $\bm{R}$ corresponding to the maximum eigenvalue $\lambda_1$. Due to the dominant eigenvector heuristic method proposed in \cite{ragi2018polynomial}, the RHS of (\ref{e2.9}) is maximized for
\begin{equation}
    \bm{\phi}_{{\rm opt}}=\exp{\{-j\phase{\bm{u}_1}\}}  \;,
    \label{e2.10}
\end{equation}
where $ \phase{\bm{u}_1}$ is a vector containing the argument of the elements of $\bm{u}_1$ and $\exp\{.\}$ is the element wise exponential operator. Here, the optimal passive beamforming $\bm{\phi}_{{\rm opt}}$ specifies the optimal phase shifts to apply at the RIS, while when substituted into (\ref{e2.6}), it also determines the optimal active beamformer to be applied at the BS. 

It is important to note that the optimal beamformers require only knowledge of the \emph{cascaded} channel $\bm{V}$, which can be estimated at the BS \cite{mishra2019channel,jensen2019optimal}, as indicated earlier. A similar optimization problem was solved in \cite{zhou2020spectral,wang2019intelligent}, however the passive beamforming vectors proposed in these works require separate knowledge of $\bm{H}$ and $\bm{h}$, which appears difficult to obtain in practice due to the passive nature of RIS.


\subsection{Maximum Received SNR}
Next, we find an expression for the maximum received SNR obtained by using the optimal beamforming vectors defined in (\ref{e2.6}) and (\ref{e2.10}). Start by expressing $\bm{R}$ as follows:
\begin{equation}
    \bm{R} = M {\rm diag}\left(\bm{h}^{*} \right) \bm{a}_{K}^{*}\left(\psi_{r}^{a},\psi_{r}^{e}\right) \bm{a}_{K}^{T}\left(\psi_{r}^{a},\psi_{r}^{e}\right) {\rm diag}\left(\bm{h} \right)
\label{e2.11}
\end{equation}
where we have used $ ||\bm{a}_{M}\left(\psi_{t}^{a},\psi_{t}^{e}\right)||^2 = M$ from (\ref{e2.3}). Since $\bm{R}$ is rank-1, we have \cite[~Prop. 1]{osnaga2005rank}
\begin{equation}
    \lambda_1= M || {\rm diag}\left(\bm{h}^{*} \right) \bm{a}_{K}^{*}\left(\psi_{r}^{a},\psi_{r}^{e}\right)||^2= M \sum_{i=1}^{K} |h_i|^2 \;,
    \label{e2.12}
\end{equation}
where we have used  $|\bm{a}_{K}^{*}\left(\psi_{r}^{a},\psi_{r}^{e}\right)_i| =1, \; \forall \; i=1,\ldots,K$. Further, the eigenvector corresponding to $\lambda_1$ can be expressed as \cite[~Prop. 1]{osnaga2005rank}
\begin{equation}
    \bm{u}_1 = \frac{{\rm diag}\left(\bm{h}^{*} \right) \bm{a}_{K}^{*}\left(\psi_{r}^{a},\psi_{r}^{e}\right)}{||{\rm diag}\left(\bm{h}^{*} \right) \bm{a}_{K}^{*}\left(\psi_{r}^{a},\psi_{r}^{e}\right)||}  = \frac{\bm{h}^{*} \odot \bm{a}_{K}^{*}\left(\psi_{r}^{a},\psi_{r}^{e}\right)}{||\bm{h}^{*} \odot \bm{a}_{K}^{*}\left(\psi_{r}^{a},\psi_{r}^{e}\right)||} 
    \label{e2.13}
\end{equation}
where $\odot $ denotes the Hadamard product. Using (\ref{e2.10}), (\ref{e2.12}) and (\ref{e2.13}) in (\ref{e2.9}), the maximum received SNR can be expressed as
\begin{equation}
    \gamma= M\Bar{\gamma} \left(\sum_{i=1}^{K}|h_i| \right)^2 \;.
    \label{e2.14}
\end{equation}

\begin{figure*}[htp] 
\centering
\subfigure[PDF ]{%
\includegraphics[width=0.5\linewidth]{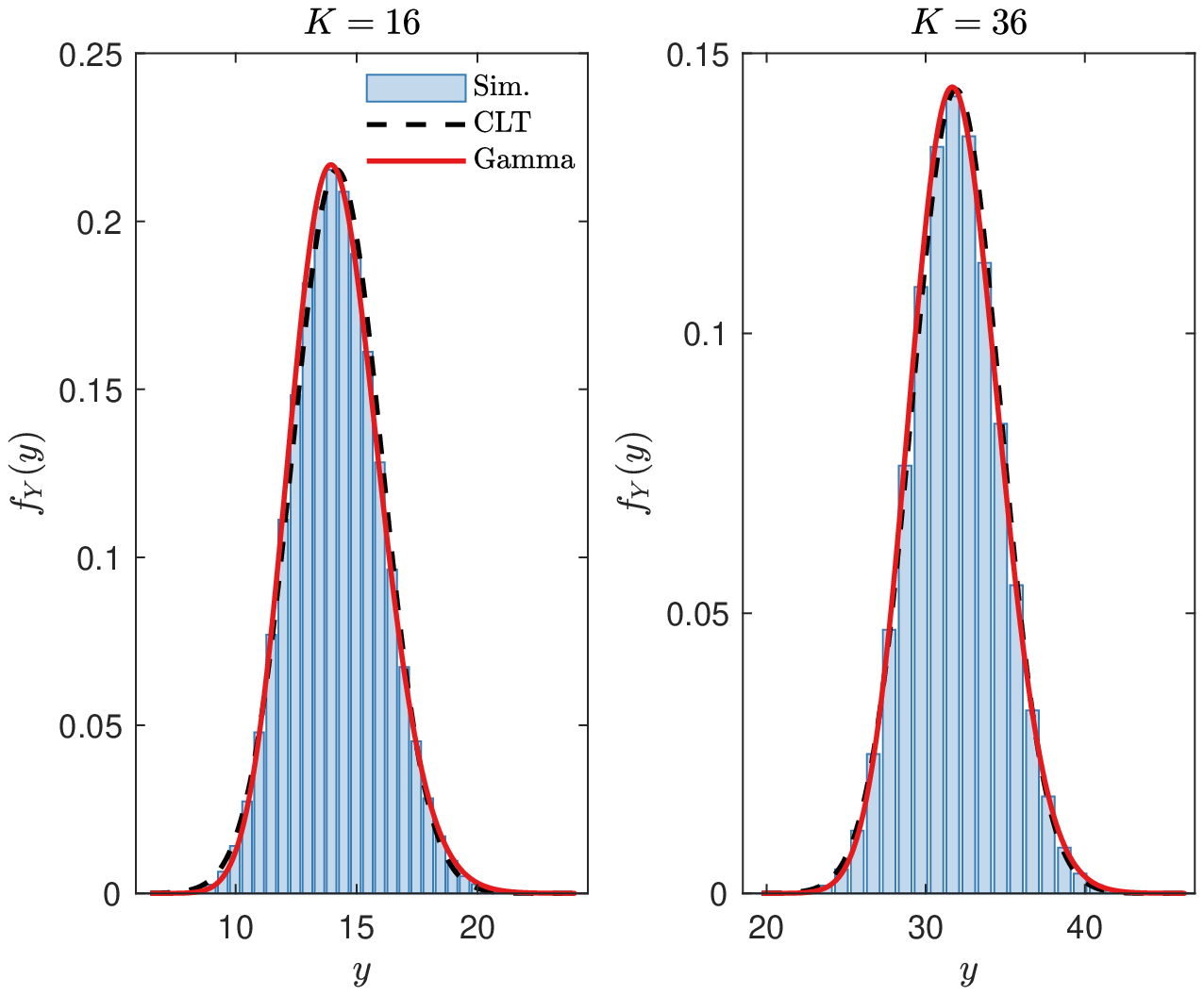}%
\label{fig3.1:a}%
}\hfil
\subfigure[CDF]{%
\includegraphics[width=0.5\linewidth]{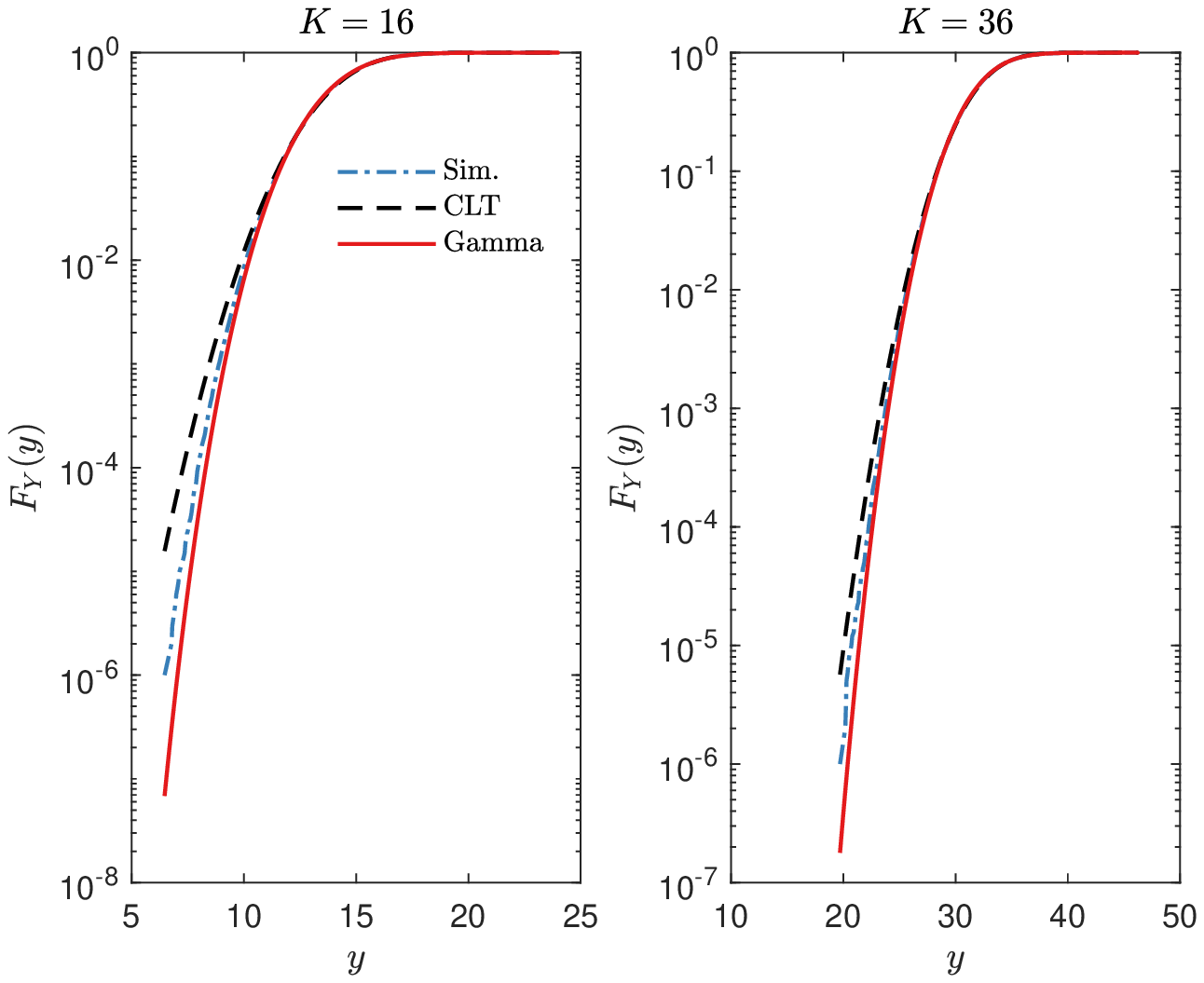}%
\label{fig3.1:b}%
}
\caption{Comparison of the simulated and theoretical distributions of $Y$, obtained from the CLT-based approximation and Gamma  approximation. Both approximations are quite accurate, with the Gamma approximation providing a better fit in the tail. }
\label{fig3.1}
\end{figure*}


\section{Statistical Characterization of $\gamma$ }

In the following, we characterize the statistical properties of the received SNR, $\gamma$.  These results will be used subsequently to study system-level performance measures. We start by characterizing the mean SNR, computed as: 
\begin{align}
    {\mathbb E}\left[ \gamma \right] 
    &= M\Bar{\gamma} \left(  \sum_{i=1}^{K}{\mathbb E}\left[|h_i|^2 \right] + \sum_{i=1}^{K}{\mathbb E}\left[ |h_i| \right] \left( \sum_{\substack{ j=1 \\ j \neq i }}^{K}{\mathbb E}\left[|h_j| \right] \right) \right) \nonumber \\
    &=\frac{M\Bar{\gamma}(K^2\pi + K(4-\pi))}{4} \;
     \label{rate2}
\end{align}
where we have used the fact that
${\mathbb E}[|h_i|]=\sqrt{\pi}/2$ and ${\mathbb E}[|h_i|^2]=1$, for all $i$, along with the independence of the $h_i$s. 
We see that the average received SNR scales quadratically with the number of RIS elements $K$, while  only linearly with the number of BS antennas $M$. Thus, the BS provides only a beamforming gain proportional to $M$, whereas the RIS provides both reflect beamforming gain proportional to $K$ and an extra aperture gain proportional to $K$ by collecting power from the BS-RIS link and then reflecting it towards the UE \cite{wu2019intelligent}.

Next, we turn to the distribution of $\gamma = M \bar{\gamma} Y^2$, where $Y= \sum_{i=1}^{K} |h_i| $, a sum of $K$ independent Rayleigh variables.  Exactly describing the distribution of $Y$, and thus $\gamma$, is challenging, and therefore one must resort to approximations \cite{hu2005accurate}.  
Here we present two such approximations: an asymptotic approximation based on the central limit theorem (CLT), and a
Gamma approximation (e.g., \cite{atapattu2011mixture}). These will be used to approximate the distribution of $\gamma$.

\emph{CLT-Based Approximation}: 
For sufficiently large $K$, the cumulative distribution function (CDF) of $Y$ can be approximated by
\begin{equation}
\begin{split}
  F_{Y}^{{\rm CLT}}(y) &=
    \left \{ \begin{array}{ll} 
    1-CQ((y-\mu_Y)/\sigma_Y) &, \;{\rm for }\, y \geq 0 \\
    0 &, \;{\rm for }\, y < 0
    \end{array} \right.
\end{split}
    \label{e3.2}
\end{equation}
and the probability density function (PDF) by
\begin{equation}
\begin{split}
    f_{Y}^{{\rm CLT}}(y) &=
    \left \{ \begin{array}{ll}  \frac{C}{\sqrt{2\pi\sigma_{Y}^2}} e^{-\frac{(y-\mu_Y)^2}{2\sigma_{Y}^2}} &, \;{\rm for }\, y \geq 0 \\
    0 &, \;{\rm for }\, y < 0
    \end{array} \right.
    \end{split}
\end{equation}
where 
\begin{align}
 \mu_{Y}= K\sqrt{\pi}/2  , \quad \sigma_{Y}^{2} = K(4-\pi)/4, 
\end{align}
and $C=1/Q(-\mu_Y/\sigma_Y)$, and where $Q(\cdot)$ is the Gaussian $Q-$function \cite{salehi2007digital}.
This follows from the CLT \cite{papoulis2002probability}, upon recognizing that $Y$ is a sum of $K$ independent, identically distributed random variables with mean ${\mathbb E}[|h_i|]=\sqrt{\pi}/2$ and variance ${\rm var}(|h_i|)=(4-\pi)/4$. Moreover, since $Y$ must necessarily be positive, we have truncated the  Gaussian distribution at zero; though, practically, for large $K$ this is inconsequential.  

\emph{Gamma Approximation:}
For the second approximation, we fit a Gamma distribution for $Y$.  This gives a CDF of the form
\begin{equation}
     F_{Y}^{\Gamma}(y) = \frac{1}{\Gamma\left(l \right)} \gamma\left(l, \frac{y}{\theta} \right) \;,
     \label{e3b.3}
\end{equation}
and PDF
\begin{equation}
    f_{Y}^{\Gamma}(y) = \frac{y^{l-1} e^{-\frac{y}{\theta}}}{\theta^{l}\Gamma\left(l \right)} \;,
\end{equation}
where 
\begin{equation}
    l= \frac{K\pi}{4-\pi}, \; \, \; \, \theta=\frac{4-\pi}{2\sqrt{\pi}} \; .
    \label{3b.2}
\end{equation}
Here, $\gamma(s,x)= \int_{0}^{x} t^{s-1} e^{-t} dt $ is the lower incomplete Gamma function, and $\Gamma(x)= \int_{0}^{\infty} t^{x-1} e^{-t} dt$ is the Gamma function. 
The Gamma distribution has two parameters: a) a shape parameter $l$, and (b) a scale parameter $\theta$, such that the mean and variance are $l\theta $ and $l\theta^2$ respectively \cite{papoulis2002probability}. The distribution (\ref{e3b.3}) follows by matching the $\mu_{Y}$ and $\sigma^{2}_{Y}$ of $Y$ with the mean $l\theta$  and variance $l\theta^2$ of the Gamma distribution. 

The two distribution approximations for $Y$ are shown in Fig. \ref{fig3.1} for $K = 16$ and $K = 36$, and compared with the exact distribution of $Y$ computed using Monte-Carlo simulations. Both approximations are found to be quite accurate, though the Gamma distribution is comparably better in the tails. 

It is now straightforward to derive distribution approximations for $\gamma$ in (\ref{e2.14}).  Recalling $\gamma= M\Bar{\gamma}Y^2$, for the CLT-based approximation, we have
\begin{align}
     F_{\gamma}^{{\rm CLT}}(z) & = {\rm Pr}(\gamma\leq z) \nonumber \\
     & = {\rm Pr}\left(-\sqrt{\frac{z}{M\Bar{\gamma}}}\leq Y \leq \sqrt{\frac{z}{M\Bar{\gamma}}}\right) \nonumber \\
     & = 
    F_{Y}^{{\rm CLT}}\left( \sqrt{\frac{z}{M\Bar{\gamma}}}\right)- F_{Y}^{{\rm CLT}}\left(- \sqrt{\frac{z}{M\Bar{\gamma}}}\right) 
     \label{e3.4}
\end{align}
for $z > 0$, and $F_{\gamma}^{{\rm CLT}}(z)=0$ otherwise. Plugging in (\ref{e3.2}) then gives the desired approximation
\begin{equation}
    F_{\gamma}^{{\rm CLT}}(z) =
    \left \{ \begin{array}{ll} 
   1-CQ\left(\frac{\sqrt{z/(M\Bar{\gamma})}-\mu_Y}{\sigma_Y} \right) &, \; {\rm for }\;  z\geq 0 \\
    0 &, \; {\rm for }\; z < 0
    \end{array} \right. .
    \label{e3.3}
\end{equation}




For the Gamma approximation, applying the same steps, but substituting (\ref{e3b.3}) rather than (\ref{e3.2}), yields the approximation 
\begin{equation}
    F_{\gamma}^{{\rm \Gamma}}(z) =
    \frac{1}{\Gamma\left(l \right)} \gamma\left(l, \frac{1}{\theta} \sqrt{\frac{y}{M\Bar{\gamma}}} \right)
    \label{ea3.5} \; .
\end{equation}

\section{Performance Analysis}

In this section, we apply the statistical properties of $\gamma$, presented above, to analyze three performance measures of the proposed RIS-MISO system: the outage probability, the average achievable rate, and the average symbol error probability for a class of digital modulation schemes.

\subsection{Outage Probability}

The distribution approximations of the received SNR $\gamma$ can be applied directly to approximate the outage probability:
\begin{equation}
    P_{{\rm out}}(\gamma_{{\rm th}})=  {\rm Pr} (\gamma \leq \gamma_{{\rm th}}) = F_{\gamma}(\gamma_{{\rm th}})\; .
    \label{e3.5}
\end{equation}
Here, one can replace the CDF of $\gamma$, $F_{\gamma}(\cdot)$, with the approximation in (\ref{e3.3}) or (\ref{ea3.5}). We will show in our numerical results in Section \ref{sec:NumRes} that these approximations are quite accurate for moderate outage levels.

It is also of interest to capture the asymptotic behavior of the outage probability as $\bar{\gamma}$ grows large, in order to quantify the diversity order achieved by the system.  For this purpose, rather than studying the approximations given above (which may be not be sufficiently accurate in the distribution tails), 
we instead apply a small argument approximation of $F_{Y}(y)$ given in \cite{brennan1959linear,hu2005accurate} as
\begin{equation}
    F_{Y}(y) \approx 1- e^{-\frac{y^2}{2dK}} \sum_{i=0}^{K-1} \frac{\left(\frac{y^2}{2dK} \right)^i}{i!}  \;,
    \label{e3b.5}
\end{equation}
where $d= \frac{\left((2K-1)!! \right)^{1/K}}{2K}$, and $(2K-1)!! = (2K-1)(2K-3)\ldots3\cdot1$. This approximation is arbitrarily accurate as $y \to 0$ \cite{brennan1959linear}.
Using the Taylor series expansion for $e^{-\frac{y^2}{2dK}}$ around $y=0$, and keeping the leading order terms, we further obtain
\begin{equation}
     F_{Y}(y)  \approx \frac{y^{2K} }{(2dK)^K K!} \; . 
     \label{e3b.7}
\end{equation}
From this, since $\gamma= M\Bar{\gamma}Y^2$, it follows that when $\gamma_{{\rm th}}/\Bar{\gamma} \rightarrow 0$, 
\begin{equation}
    P_{{\rm out}}(\gamma_{\rm th}) \approx \frac{1}{M^K (2K-1)!!} \left(\frac{\gamma_{{\rm th}}}{\Bar{\gamma}} \right)^K \, .
    \label{e3b.4}
\end{equation}






As is well-known \cite{wang2003simple}, a high SNR outage probability approximation of the form $P_{{\rm out}} \approx \left(O_c \Bar{\gamma} \right)^{-G_d}$, implies a diversity gain of $G_d$ and a coding (or array) gain of $O_c$. Hence, for the RIS-MISO system, a diversity order of  $K$ is achieved, which scales linearly with the number of RIS elements, but has no dependence on the number of BS antennas $M$.  This is consistent with a result shown in \cite{kudathanthirige2020performance}, for a RIS-SISO system. The coding gain $O_c=(M/\gamma_{{\rm th}})((2K-1)!!)^{1/K} $, on the other hand, depends on both $K$ and $M$, and notably, grows linearly with $M$. That is, increased diversity order (i.e., an increased effective number of independent channels) can be achieved by increasing the number of RIS elements, whereas increasing the number of BS antennas enhances performance by offering an effective power gain through active beamforming at the BS.

\subsection{Average Achievable Rate}
The achievable rate at the UE is given by
\begin{equation}
    R={\mathbb E}\left[ \log_2\left(1+ \gamma \right) \right] \; .
    \label{e3.6}
\end{equation}
Using Jensen's inequality, this is upper bounded as
\begin{equation}
    R\leq \log_{2}\left(1+ {\mathbb E}\left[ \gamma \right] \right)= R_{{\rm ub}} \; 
    \label{e3.8}
\end{equation}
which, upon substituting (\ref{rate2}), gives 
\begin{equation}
    R_{{\rm ub}} = \log_{2}\left(1+ \frac{M\Bar{\gamma}(K^2\pi + K(4-\pi))}{4} 
    \right) \; .
    \label{e3.7}
\end{equation}
For large $K$, we see that $R_{{\rm ub}} \sim  \log_2 \left( \bar{\gamma} M K^2 \pi / 4 \right)$, indicating that while the RIS elements provide linear growth in diversity gain as well as a power boost (i.e., reflected by the coding gain), they provide no additional benefit in terms of multiplexing gain, as one may expect.



\subsection{Average Symbol Error Probability }
Finally, we analyse the average symbol error probability (SEP) for uncoded digital modulation schemes.  For numerous modulation schemes (e.g., BPSK, QPSK), the SEP, conditioned on $\gamma$, can be expressed as $P_{e|\gamma}(\bar{\gamma}) = \alpha Q\left(\sqrt{\beta \gamma} \right) $, where $\alpha, \beta$ are  modulation specific parameters \cite{salehi2007digital}. Since $\gamma=M\Bar{\gamma} Y^2$, the average SEP $\Bar{P_{e}}(\bar{\gamma}) = {\mathbb E}\left[ \alpha Q\left(\sqrt{\beta \gamma} \right) \right]$ is given by 
\begin{align}
    \Bar{P_e}(\bar{\gamma}) &= \int_{0}^{\infty}  \alpha Q\left(\sqrt{M\beta \Bar{\gamma}}y \right) f_{Y}(y) dy \;.
    \label{e3.13}
\end{align}
To evaluate this, we use the CLT approximation for $f_{Y}(y)$ in (\ref{e3.13}) which, after some algebraic manipulations, yields 
\begin{equation}
    \Bar{P_e}(\bar{\gamma}) = \Upsilon \int_{0}^{\infty} \exp{\left(-(by^2-2cy)\right)}Q\left(y\sqrt{a} \right) dy
    \label{e3.15}
\end{equation}
where $\Upsilon, a,b$, and $c$ are constants defined as
\begin{equation}
   \Upsilon=\frac{\alpha C \exp{\left(-\frac{\mu_{Y}^{2}}{2\sigma_{Y}^{2}}\right)}  } {\sqrt{2\pi} \sigma_Y}, \,  a=M\beta\Bar{\gamma}, \, b= \frac{1}{2\sigma_{Y}^{2}},\, c= \frac{\mu_Y}{2\sigma_{Y}^2}.
   \label{e3.11}
\end{equation}
Next, it is convenient to use the $Q(x)$-function representation \cite{gradshteyn} 
\begin{equation}
    Q(x) = \frac{1}{\pi} \int_{0}^{\frac{\pi}{2}} \exp{\left(-\frac{x^2}{2\sin^2{\theta}}  \right)} d\theta \;,
    \label{e3.16}
\end{equation}
in (\ref{e3.15}) which, after some manipulations, leads to 
\begin{equation}
    \Bar{P_e}(\bar{\gamma}) = \frac{\Upsilon}{\pi}  \int_{0}^{\frac{\pi}{2}} \int_{0}^{\infty} \exp{\left( -\left(\frac{a}{2\sin^2{\theta}}+b \right)y^2 + 2cy \right)} dy d\theta.
    \label{e3.17}
\end{equation}
Simplifying the inner integral by using \cite[eq.~2.33.1]{gradshteyn}, we obtain 
\begin{equation}
\begin{split}
   \Bar{P_e}(\bar{\gamma}) = \frac{\Upsilon}{\sqrt{\pi}} \int_{0}^{\frac{\pi}{2}} \frac{\exp{\left(\frac{c^2}{\frac{a}{2\sin^2{\theta}}+b} \right)}}{\sqrt{\frac{a}{2\sin^2{\theta}} + b}}  Q\left(\frac{-\sqrt{2}c}{\sqrt{\frac{a}{2\sin^2{\theta}} + b}} \right) d\theta\;.
   \label{e3.18}
   \end{split}
\end{equation}
It appears difficult to solve this explicitly, though, due to the finite integration limits, it can be easily evaluated with numerical integration. Moreover, an analytical upper bound can be obtained by setting $\theta=\pi/2$ \cite{kudathanthirige2020performance}, which gives
\begin{equation}
 \Bar{P_{e}}^{{\rm ub}}(\bar{\gamma}) =\frac{\Upsilon \sqrt{\pi}}{\sqrt{ 2a +4b }} \exp{ \left(\frac{c^2}{\frac{a}{2}+b}\right)} Q\left(\frac{-\sqrt{2}c}{\sqrt{\frac{a}{2}+b}} \right) \;.
 \label{e3.10}
\end{equation}

\section{Numerical Results} \label{sec:NumRes}
We consider three scenarios with different $M$ and $K$: Case-1: $M=16, K=16$,  Case-2: $M=36, K=16$, and Case-3: $M=16, K=36$. For the array response vectors at the BS and RIS we assume $\psi_{r}^{a} = \psi_{r}^{e}=\psi_{t}^{a} = \psi_{t}^{e} = \pi/4 $, and $d/\lambda = 0.5$.
\begin{figure}[h] 
\centering
\includegraphics[width=0.5\textwidth]{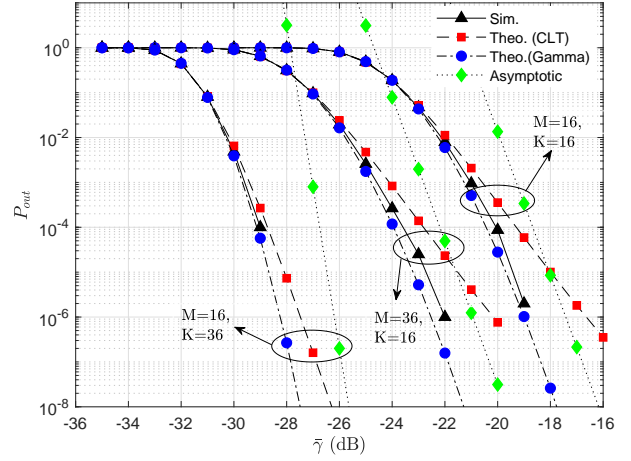}%
\caption{The plots compare the outage probability $P_{{\rm out}}(\bar{\gamma})$ obtained from Monte-Carlo simulations, the theoretical $P_{{\rm out}}(\bar{\gamma})$ obtained from the CLT approximation (\ref{e3.3}) and Gamma distribution approximation (\ref{ea3.5}), and the asymptotic $P_{{\rm out}}(\bar{\gamma})$ from (\ref{e3b.4}), with fixed $\gamma_{{\rm th}} = 10$ dB. }
\label{fig4.1}
\end{figure}

Fig. \ref{fig4.1} compares the outage probability $P_{{\rm out}}(\bar{\gamma})$ obtained from Monte-Carlo simulations, the theoretical $P_{{\rm out}}(\bar{\gamma})$ using the approximate CDF expression of $\gamma$ obtained from the CLT (\ref{e3.3}) and Gamma distribution approximation (\ref{ea3.5}), and the asymptotic $P_{{\rm out}}(\bar{\gamma})$ from (\ref{e3b.4}). The theoretical $P_{{\rm out}}(\bar{\gamma})$ obtained from both the CLT and Gamma approximations are fairly accurate for moderate outage levels, though at low outages the Gamma approximation is more accurate. From the slope of the asymptotic $P_{{\rm out}}$, it can be observed that when $M$ increases with fixed $K$ (Case-1 and Case-2), the diversity order remains the same, whereas it increases in Case-3 due to the larger $K$.  Since Case-3 achieves around $9$ dB gain compared to Case-1, whereas Case-2 achieves only $4$ dB gain, we conclude that increasing $K$ leads to higher performance improvement compared to increasing $M$.

\begin{figure}[h] 
\centering
\includegraphics[width=0.5\textwidth]{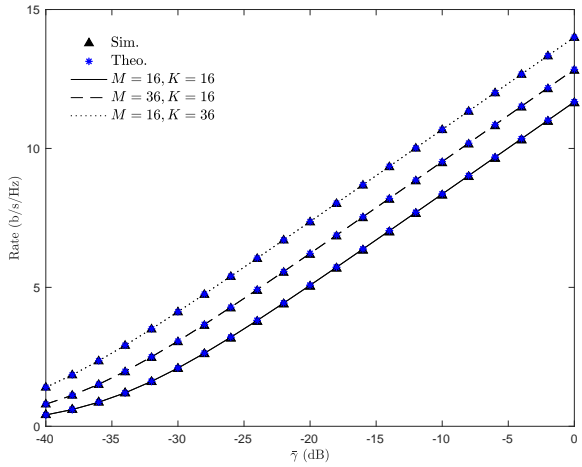}%
\caption{The plots show the achievable rate obtained from Monte-Carlo simulations and the theoretical upper bound given by (\ref{e3.7}) as the average transmit SNR increases.}
\label{fig4.2}
\end{figure}

Average achievable rates are shown in Fig. \ref{fig4.2}. Again, the gain due to increasing $K$ (Case-3) is higher than that due to increasing $M$ (Case-2), since the average received SNR scales as $K^2$, but only linearly in $M$. The slope of the rate curves, and hence the multiplexing gain, is the same in all cases. 

\begin{figure}[h] 
\centering
\includegraphics[width=0.5\textwidth]{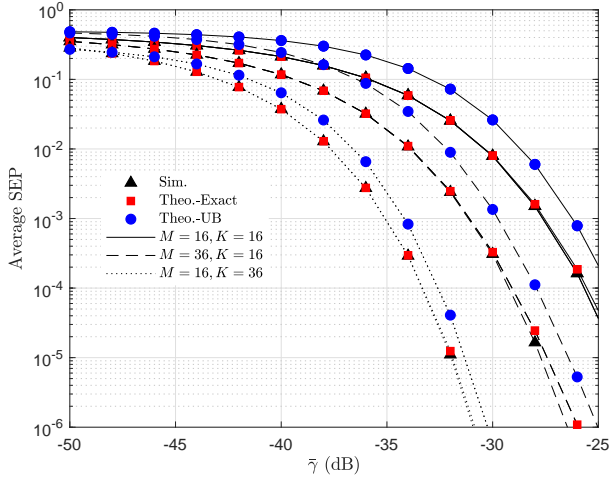}%
\caption{The plots compare the average SEP of BPSK modulation obtained from Monte-Carlo simulations, the exact theoretical SEP obtained by numerical integration of (\ref{e3.18}), and the upper bound given by (\ref{e3.10}). }
\label{fig4.3}
\end{figure}

Fig. \ref{fig4.3} shows the average SEP for BPSK, for which $\alpha=1, \; {\rm and}\; \beta=2$. The plot shows the average SEP obtained from Monte-Carlo simulations, the exact theoretical SEP obtained from numerical integration of (\ref{e3.18}), and the theoretical upper bound given by (\ref{e3.10}). The results validate the exact analysis, while confirming the validity of the upper bound. Here, Case-3 achieves around $8$ dB gain compared to Case-1, whereas Case-2 achieves around $3$ dB gain.

\section{Conclusion}
We considered a RIS-assisted MISO system, where the BS-RIS link is LoS, and the RIS-UE link experiences Rayleigh fading. We presented closed-form expressions for the optimal beamforming vectors, along with analysis of outage probability, achievable rate and SEP. Our analysis reveals that the diversity order is equal to the number of RIS elements $K$, with no dependence on the number of BS antennas $M$, while the coding gain depends on both $K$ and $M$. The average received SNR scales linearly with $M$, and  quadratically with $K$.  

The optimal beamformers will change if the BS-RIS channel is subjected to fading, rather than LoS as assumed in our system model. This will also require new performance analysis. Computing the optimal beamformers and system performance under conditions where both the BS-RIS and RIS-UE links exhibit Rayleigh fading appears challenging, and this remains a problem to be addressed in future research.

\bibliographystyle{IEEEtran}
\bibliography{IEEEabrv,LIS_MISO}


\end{document}